\documentclass[a4paper]{PoS}
\pdfoutput=1
\usepackage{amsmath,amssymb,amsbsy,multicol,url}
\usepackage[utf8]{inputenc}
\usepackage{mathtools}
\usepackage[numbers,sort&compress]{natbib}
\usepackage{natbibspacing}
\usepackage{graphicx}
\usepackage{ulem,color}
\title{Non-perturbative determinations of $B$-meson decay constants and semi-leptonic form factors}

\ShortTitle{Non-perturbative determinations of $B$-meson decay constants and semi-leptonic form factors}

\author{Edwin Lizarazo$^a$ and \speaker{Oliver Witzel}$^b$\newline
        \textbf{(RBC and UKQCD collaborations)}\\
        $^a$ Physics and Astronomy, University of Southampton, Southampton, UK\\
        $^b$ Higgs Centre for Theoretical Physics, School of Physics and Astronomy\\University of Edinburgh, Edinburgh, UK\\
        E-mail: \email{e.lizarazo@soton.ac.uk, o.witzel@ed.ac.uk}}

\abstract{$B$-physics is one of the most promising windows to find new physics in the flavor sector. One key
  ingredient to these searches are precise theoretical predictions derived from the Standard
  Model. Focusing at the nonperturbative QCD contributions, we carry out lattice QCD
  simulations in order to calculate $B$-meson decay constants and semi-leptonic form factors.
  Combined with experimental measurements our results enable us to determine CKM matrix
  elements.\\

  Here we present $B$ and $B_s$ meson decay constants as well as semi-leptonic form factors
  including rare decays, CKM or GIM suppressed in the Standard Model. Our results are
  based on the set of 2+1 flavor domain-wall Iwasaki gauge field configurations generated by
  the RBC-UKQCD collaboration. Heavy $b$-quarks are simulated using the relativistic heavy
  quark action.}

\FullConference{38th International Conference on High Energy Physics\\
		3-10 August 2016\\
		Chicago, USA}

\begin{document}
\section{Introduction}
$B$-physics plays a central role in fits of the CKM unitarity triangle and helps by that to derive constraints on new physics. While experiments like BaBar, Belle, LHCb, Atlas, CMS, and, in the future, Belle II carry out precise measurements of various $B$-meson observables, we also need theoretical predictions derived from the Standard Model (SM) to compare to or isolate fundamental parameters like CKM matrix elements. At short distance the interactions are dominated by the strong force and require non-perturbative methods to calculate the contribution from the SM. Here we focus on such non-perturbative $B$-physics computations using lattice QCD.

\begin{figure}[b]
  \centering
\parbox{0.34\textwidth}{
\begin{picture}(50,27)(25,49)
    \put(30,52){\includegraphics[width=40mm]{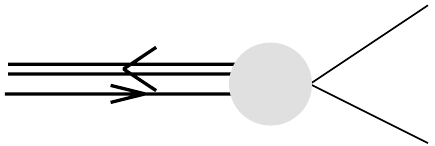}}
    \put(25,75){a)}
    \put(26,57){\large{$B$}}
    \put(70,63){\large{$\ell$}}
    \put(70,50){\large{$\nu_\ell$}}
\end{picture}}
\hspace{10mm}
\parbox{0.42\textwidth}{
\begin{picture}(63,27)(-10,24)
    \put(-4,25){\includegraphics[width=50mm]{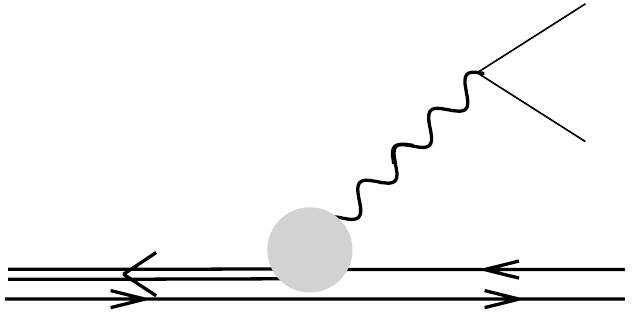}}
    \put(-8,50){b)}    
    \put(-10,26){\large{$B_{(s)}$}} \put(47,26){\large{$\pi\, (K)$}}
    \put(22,36){\large{$W$}} 
    \put(43,48){\large{$\ell$}} \put(43,37){\large{$\nu_\ell$}}
\end{picture}}\\[5mm]
\parbox{0.42\textwidth}{
\begin{picture}(63,27)(52,24)
    \put(60,25){\includegraphics[width=50mm]{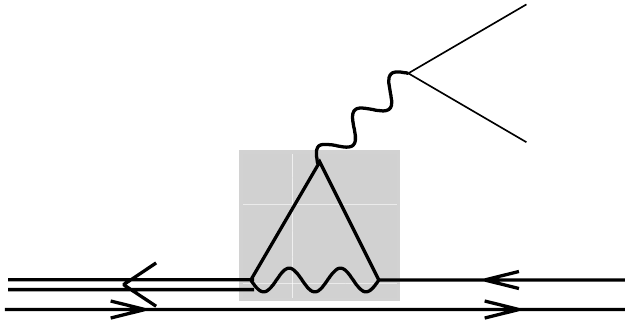}}
    \put(58,49){c)}     
    \put(55,26){\large{$B_s$}} \put(111,26){\large{$\phi$}}
    \put(80,33){\large{$t$}} \put(89,33){\large{$t$}} 
    \put(82.5,30){\large{$W$}} \put(79,42){\large{$Z,\gamma$}} 
    \put(102,49){\large{$\ell$}} \put(102,38){\large{$\ell$}}
\end{picture}}
\hspace{10mm}
\parbox{0.42\textwidth}{
\begin{picture}(63,28)(55,42)
    \put(60,44){\includegraphics[width=50mm]{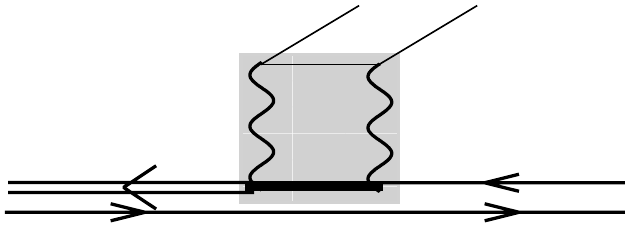}}
    \put(51,67){d)}        
    \put(55,45){\large{$B_s$}} \put(110,45){\large{$\phi$}}
    \put(75,51){\large{$W$}} \put(92,51){\large{$W$}}
    \put(89,63){\large{$\ell$}} \put(98.5,63){\large{$\ell$}}
    \put(84,48.5){\large{$t$}} \put(86,57){\large{$\nu$}}
\end{picture}}
\caption{Diagrams sketching the calculation of non-perturbative, short distance contributions to leptonic and semi-leptonic decays: a) pure leptonic decay (decay constant), b) charged tree-level decay ($B\to\pi \ell \nu$ or $B_s\to K\ell\nu$), c) and d) loop-level decays with flavor changing neutral currents (e.g., $B_s\to\phi\ell^+\ell^-$). The short distance contributions are indicated by the gray shading and implemented as point-like operators.}
\label{Fig:DiagramSketches}
\end{figure}
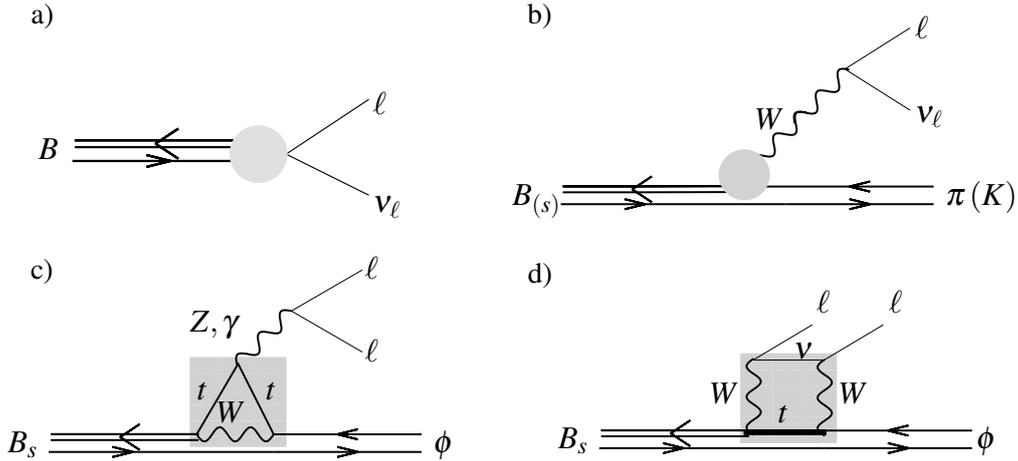

Numerical simulations of $b$-quarks face the challenge that typically available lattice QCD gauge field ensembles are generated with a cutoff around or below 3 GeV, but the mass of the $b$-quark is larger, about 4.3 GeV. Hence we simulate $b$-quarks using an effective action, the relativistic heavy quark (RHQ) \cite{Christ:2006us,Lin:2006ur} or Fermilab \cite{ElKhadra:1996mp} action which is based on the anisotropic Sheikoleslami-Wohlert action \cite{Sheikholeslami:1985ij} with a special interpretation of its three parameters such that discretization errors are small. After tuning these parameters non-perturbatively \cite{Aoki:2012xaa}, we use the same set-up for the calculation of various $B$-physics quantities of interest. Our calculations are based on RBC-UKQCD's 2+1 flavor domain-wall fermion and Iwasaki gauge action ensembles \cite{Allton:2008pn,Aoki:2010dy,Blum:2014tka}. For the light and strange quarks we also use domain-wall fermions \cite{Kaplan:1992bt,Furman:1994ky}. In the next section we present results for our calculation $B$-meson decay constant and $B\to \pi\ell\nu$ semi-leptonic form factors and will subsequently report on our ongoing work to determine semi-leptonic form factors for rare decays only occurring at loop-level in the SM. The diagrams discussed in the following are sketched in Fig.~\ref{Fig:DiagramSketches}.

\section{Results for $B$-meson decay constants and $B\to\pi\ell\nu$ semi-leptonic form factors}
\begin{figure}[tb]
  \centering
   \parbox{0.45\textwidth}{\includegraphics[height=0.24\textheight]{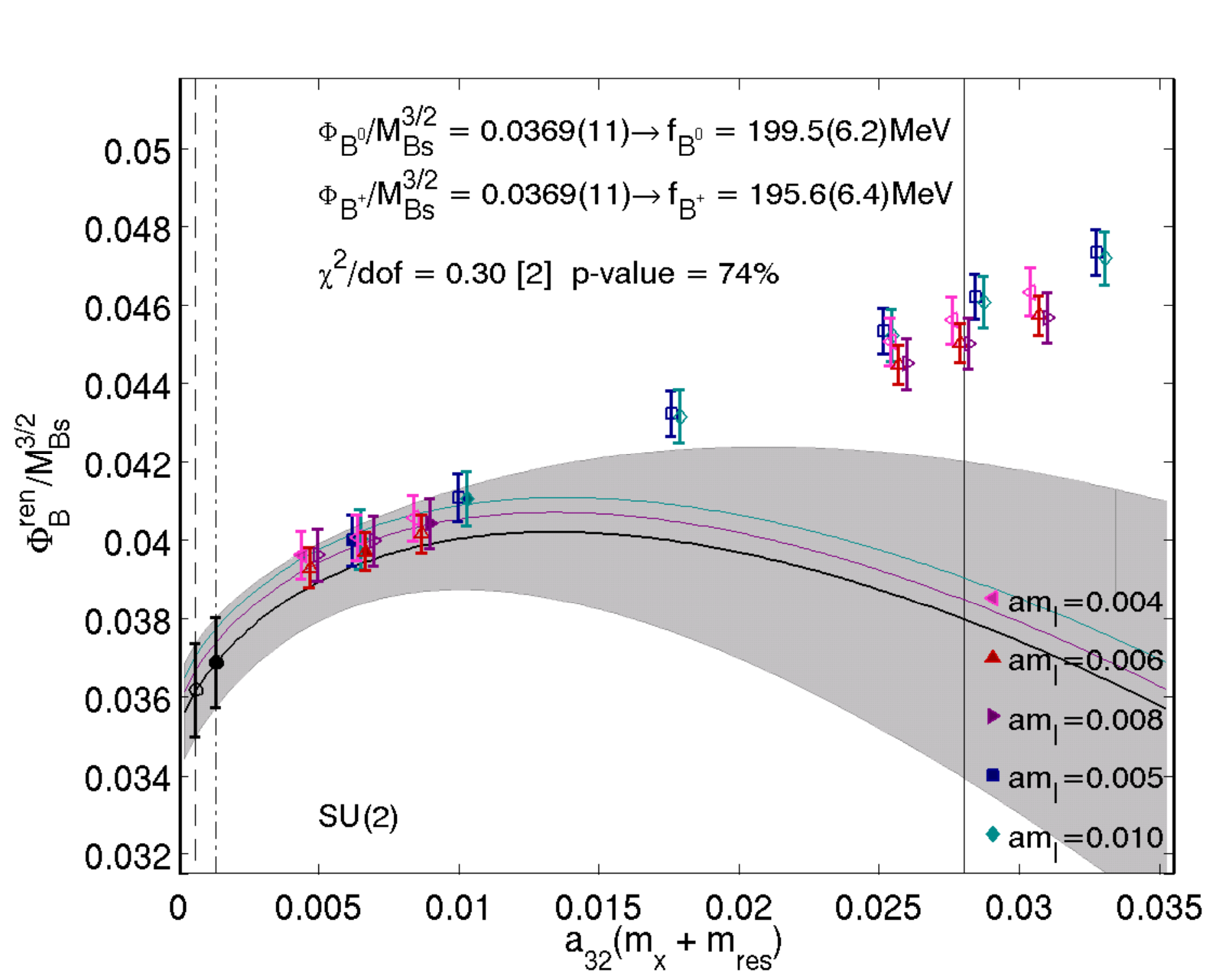}}
   \parbox{0.45\textwidth}{\includegraphics[height=0.24\textheight]{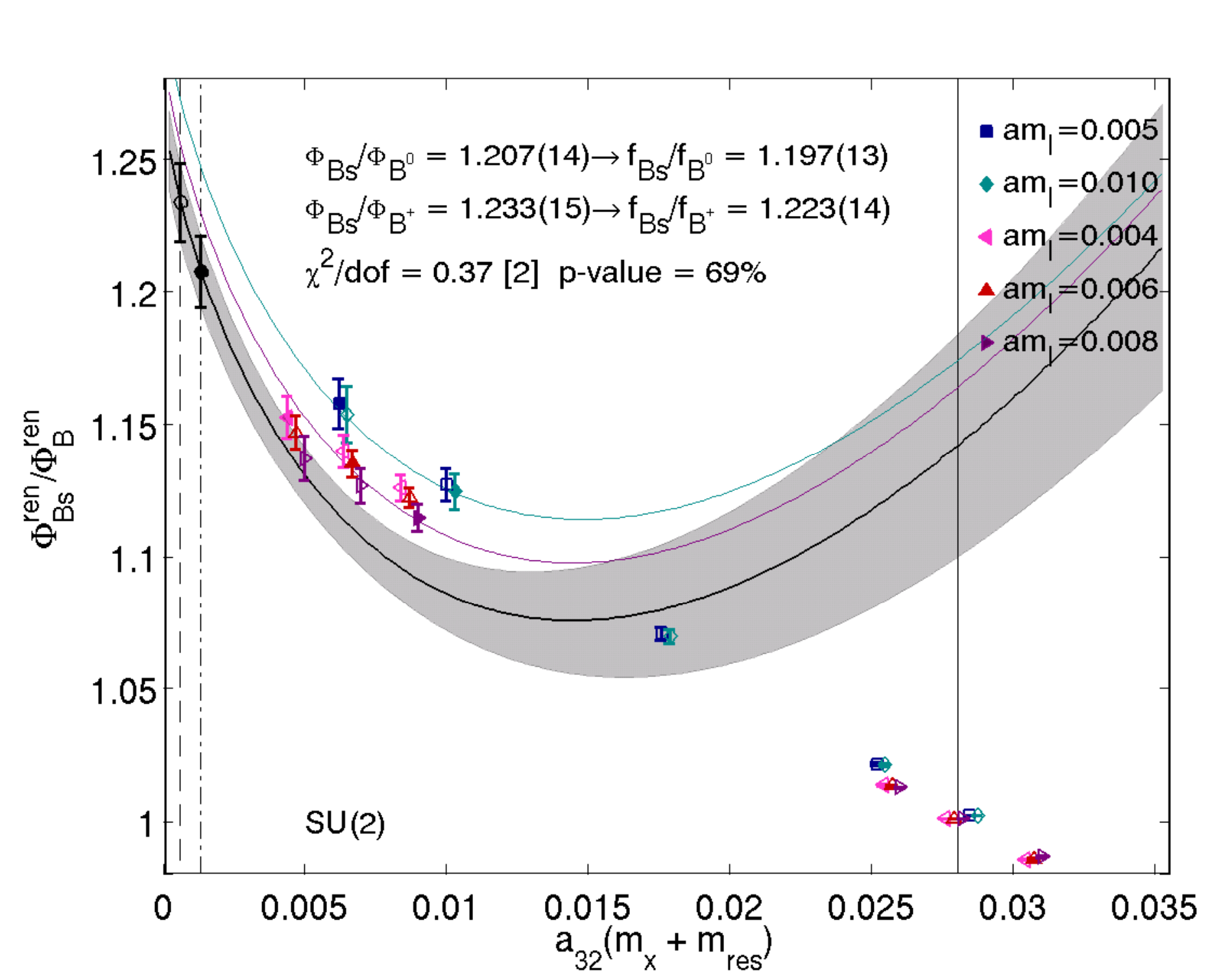}}
   \caption{Chiral and continuum extrapolation of $\Phi_{B_q}$ (left) and  $\Phi_{B_s}/\Phi_{B_q}$ (right) from a correlated fit using NLO SU(2) HM$\chi$PT.  The different colors/symbols distinguish our data points on the five different ensembles.  For better visibility data points on the $a m_l=0.004,\,0.008,\,0.01$ ensembles are plotted with a small horizontal offset.   {We plot all partially-quenched data, but the fit only includes the five unitary points (filled).}   
     The chiral extrapolation in full QCD and the continuum is shown by the black line with gray error band.  The physical values of $\Phi_{B^+}$ ($\Phi_{B^0}$) and $\Phi_{B_s}/\Phi_{B^+}$ ($\Phi_{B_s}/\Phi_{B^0}$) correspond to the intersection of this curve with the dashed (dot-dashed) vertical line on the left-hand side indicating the physical $u$-quark ($d$-quark) mass.  The right-hand solid, vertical line indicates the $s$-quark mass.  Only statistical errors are shown.}
  \label{fig:DecayConstants}
\end{figure}

The non-perturbative calculation of $B$-meson decay constants requires only the relative simple evaluation of 2-point functions as is depicted in Fig.~\ref{Fig:DiagramSketches}a indicating the creation and later annihilation of $B$-meson giving rise to a lepton-neutrino pair. We calculate and renormalize these 2-point functions for a set of five gauge field ensembles featuring two different lattice spacing and five different pion masses in the sea-sector down to 289 MeV. In the left (right) plot of Fig.~\ref{fig:DecayConstants}, we show these results for the decay amplitude $\Phi_B$ (the ratio $\Phi_{B_s}/\Phi_B$) in lattice units. The decay amplitude is proportional to the decay constant $f_B = \Phi_B/\sqrt{M_B}$ with $M_B$ the mass of the $B$-meson. 
The gray band shows the result of our combined chiral- and continuum extrapolation obtained by fitting the data points with filled symbols using SU(2) heavy meson chiral perturbation theory (HM$\chi$PT) \cite{Goity:1992tp,Arndt:2004bg,Aubin:2005aq,Albertus:2010nm}. Reading off the values at the physical quark masses indicated by the vertical lines, we obtain
\begin{align}
        f_{B^0} = {199.5(6.2)(12.6)} \;\text{MeV,}\quad  f_{B^+}  =  &{195.6(6.4)(14.9)} \;\text{MeV,}\quad f_{B_s} =  235.4(5.2)(11.1) \;\text{MeV,} \nonumber\\ 
       f_{B_s}/f_{B^0} = {1.197(13)(49),} &\quad \text{and} \quad f_{B_s}/f_{B^+}  = {1.223(14)(70)}  \label{eq:fBs/fB+} \,,    
\end{align}
where the first error indicates the statistical error and the second our combined estimate for the systematic uncertainties. For further details see Reference \cite{Christ:2014uea}.

\begin{figure}[tb]
  \centering
  \parbox{0.45\textwidth}{\includegraphics[height=0.22\textheight]{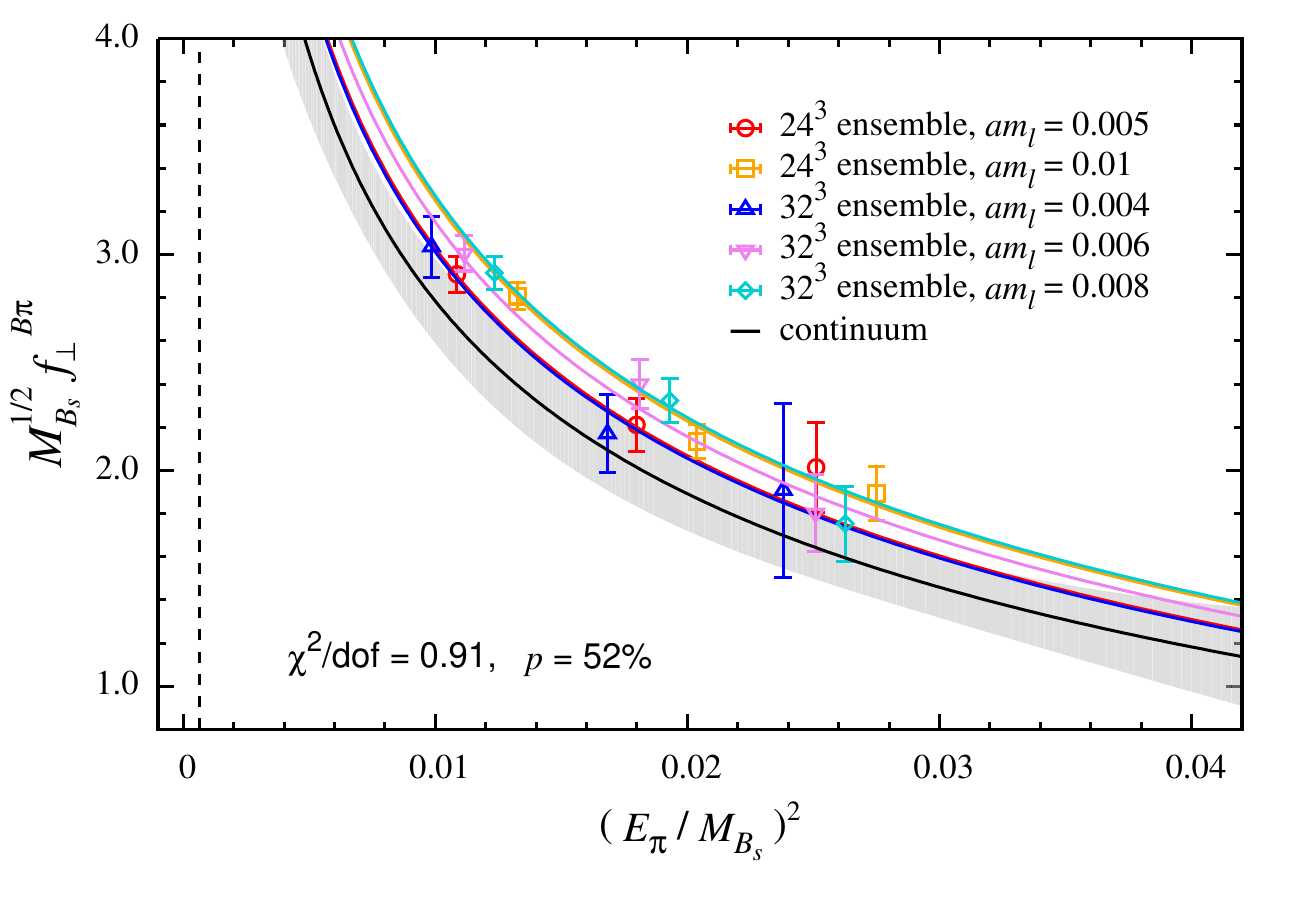}}
  \hfill
  \parbox{0.45\textwidth}{\includegraphics[height=0.22\textheight]{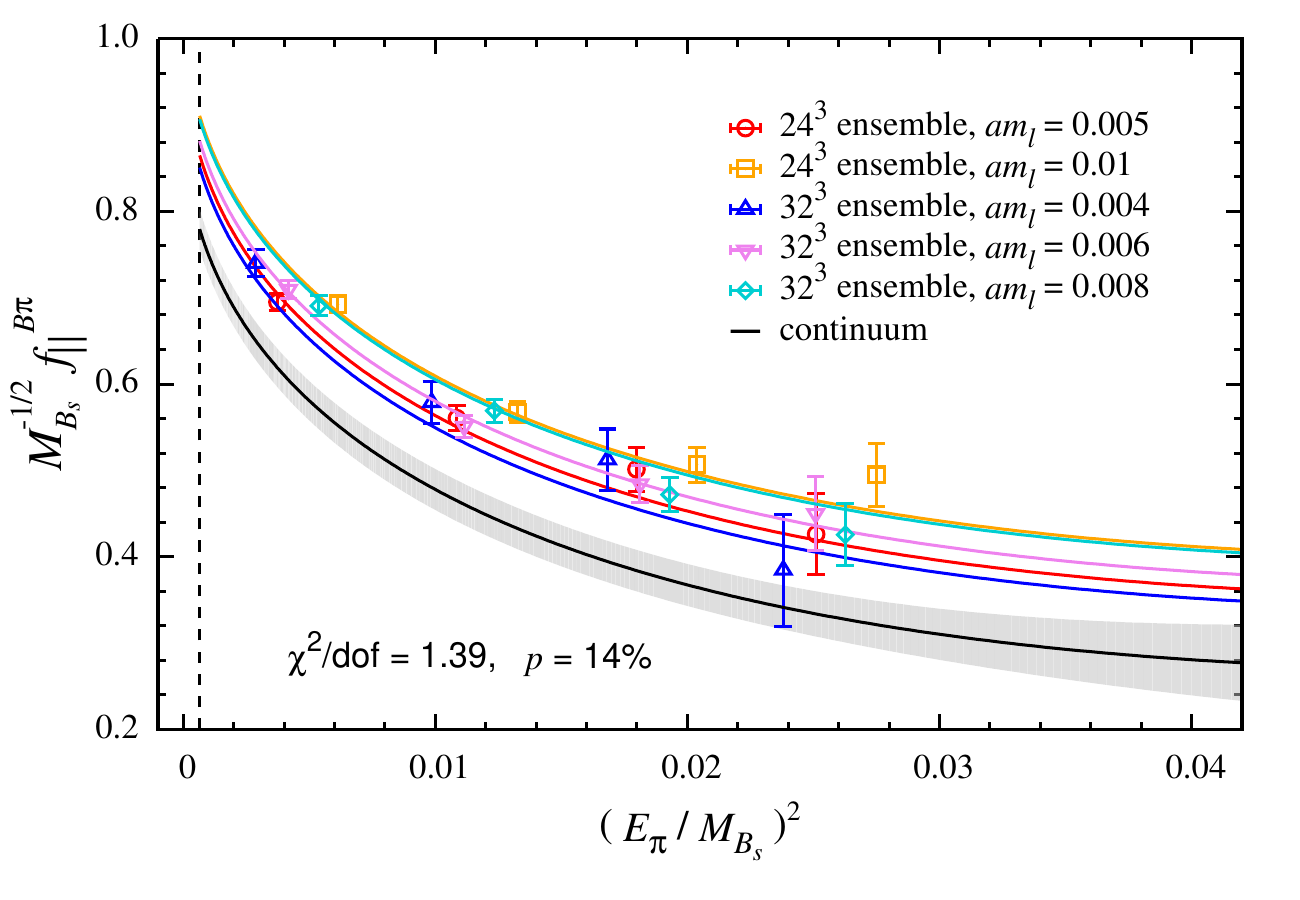}}
  \caption{Chiral-continuum extrapolation of the $B\to\pi\ell\nu$ form factors from correlated fits using NLO SU(2) hard-pion HM$\chi$PT. The colors distinguish between data points on the five different ensembles. The fit function is evaluated at the unphysical sea-quark masses and nonzero lattice spacings on the different ensembles, such that the curves should go through the data points of the same color.  The continuum, physical-quark-mass form factors are shown as a function of pion energy by the black lines with gray error band.  The vertical dashed line on the left-hand side of each plot shows the physical pion  mass.}
  \label{fig:SemiLeptonic0}  
\end{figure}

\begin{figure}[tb]
  \centering
  \parbox{0.45\textwidth}{\includegraphics[height=0.24\textheight]{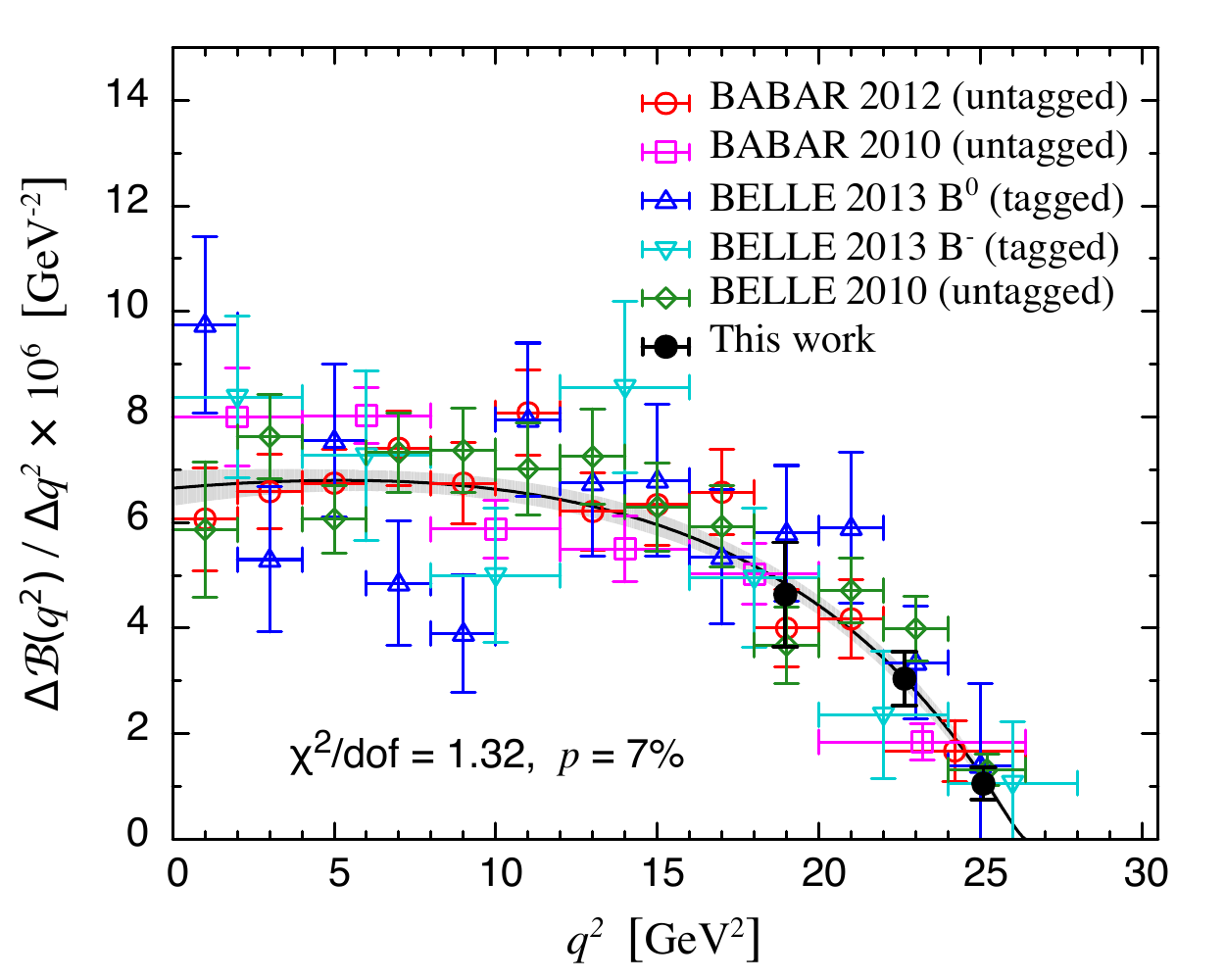}}
  \parbox{0.45\textwidth}{\includegraphics[height=0.24\textheight]{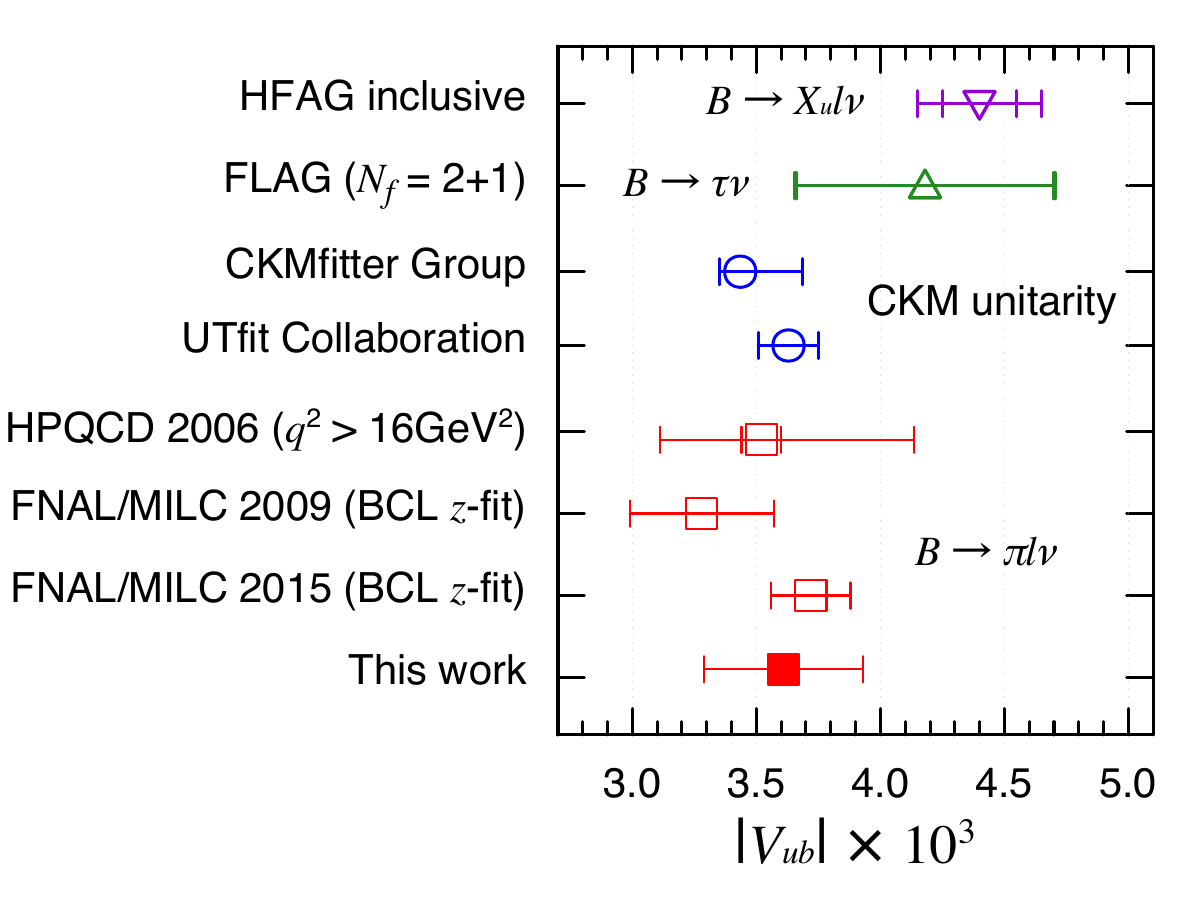}}
   \caption{Left: Model-independent determination of $|V_{ub}|$ from a combined fit of experimental measurements of the $B \to \pi\ell\nu$ branching fraction \cite{delAmoSanchez:2010af,Lees:2012vv,Ha:2010rf,Sibidanov:2013rkk} and our lattice result. Right: Comparison of $|V_{ub}|$ determinations \cite{Amhis:2012bh,Aoki:2013ldr,CKMfitterWinter2014,UTfitSummer2014,Dalgic:2006dt,Bailey:2008wp,Lattice:2015tia}. For points with double error bars, the inner error bars are experimental while the outer error bars show the total experimental plus theoretical uncertainty added in quadrature.}
  \label{fig:SemiLeptonic}
\end{figure}

Using the same set of gauge field configurations, we computed semi-leptonic decays occurring at tree-level and mediated by a charged current. Specifically we obtained the form factors for $B\to\pi\ell\nu$ and $B_s\to K\ell\nu$ decays and depict the corresponding diagram in Fig.~\ref{Fig:DiagramSketches}b. On the lattice we evaluate the non-perturbative contributions by implementing 3-point functions locating the $B_{(s)}$-meson to be at $t_\text{sink}$, the pion (kaon) to be at $t_0$, and then calculate the contributing effective operators for the time slices in between. We keep the $B_{(s)}$-meson at rest and inject momentum on the pion (kaon) side in order to explore the range of momentum transfer ($q^2$). The results of our lattice simulations for $B\to\pi\ell\nu$ are shown in Fig.~\ref{fig:SemiLeptonic0} where we show the form factors $f_\perp$ and $f_\parallel$ which are linearly related to the phenomenologically used quantities $f_+$ and $f_0$. After carrying out a combined chiral- and continuum extrapolation, we account for our systematic uncertainties and end up with three so-called synthetic data points. We show these three synthetic data points by the black symbols in the left plot of Fig.~\ref{fig:SemiLeptonic}. The colored symbols refer to the experimental measurements by BaBar \cite{delAmoSanchez:2010af,Lees:2012vv} and Belle \cite{Ha:2010rf,Sibidanov:2013rkk}. The plot is made after fitting the normalization constants which is equal to the CKM matrix element $|V_{ub}|$. Our calculation leads to the determination of
\begin{align}
  |V_{ub}| = 3.61(32) \times 10^{-3},
\end{align}
where the error combines statistical and all systematic uncertainties. The right plot in Fig.~\ref{fig:SemiLeptonic} shows the comparison of our result to other determinations of $|V_{ub}|$. Except for the persisting $2-3 \sigma$ discrepancy to the inclusive determination our result in good agreement with other determinations \cite{Amhis:2012bh,Aoki:2013ldr,CKMfitterWinter2014,UTfitSummer2014,Dalgic:2006dt,Bailey:2008wp,Lattice:2015tia}. Further details are presented in Reference \cite{Flynn:2015mha}.

\section{Outlook: rare $B$ decays with loop-level flavor changing neutral currents}

\begin{figure}[tb]
  \vspace{-4mm}
  \begin{picture}(148,88)
  \put(0,66){    
  \parbox{0.47\textwidth}{\includegraphics[height=0.19\textheight]{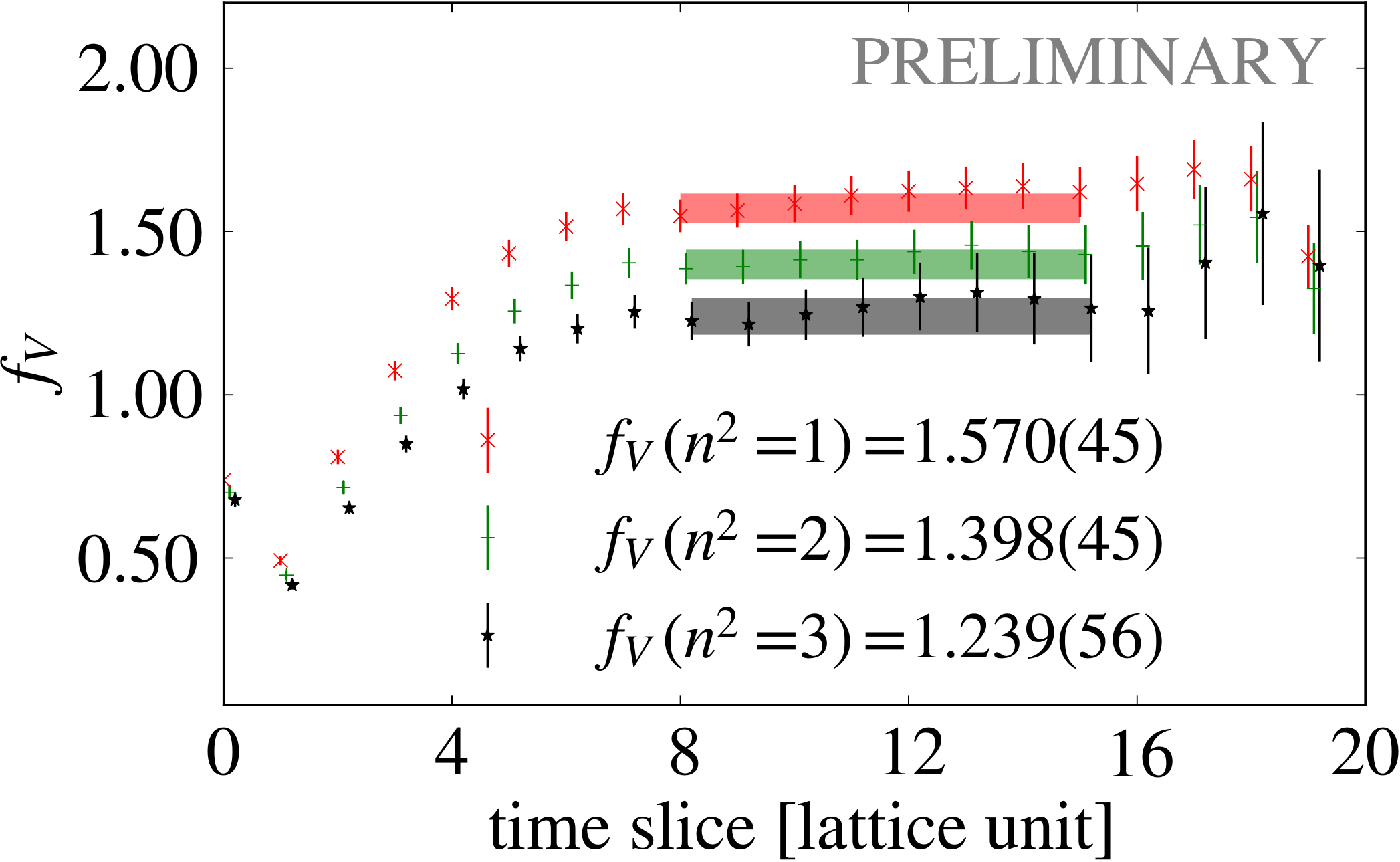}}\hspace{8.1mm}
  \parbox{0.47\textwidth}{\includegraphics[height=0.19\textheight]{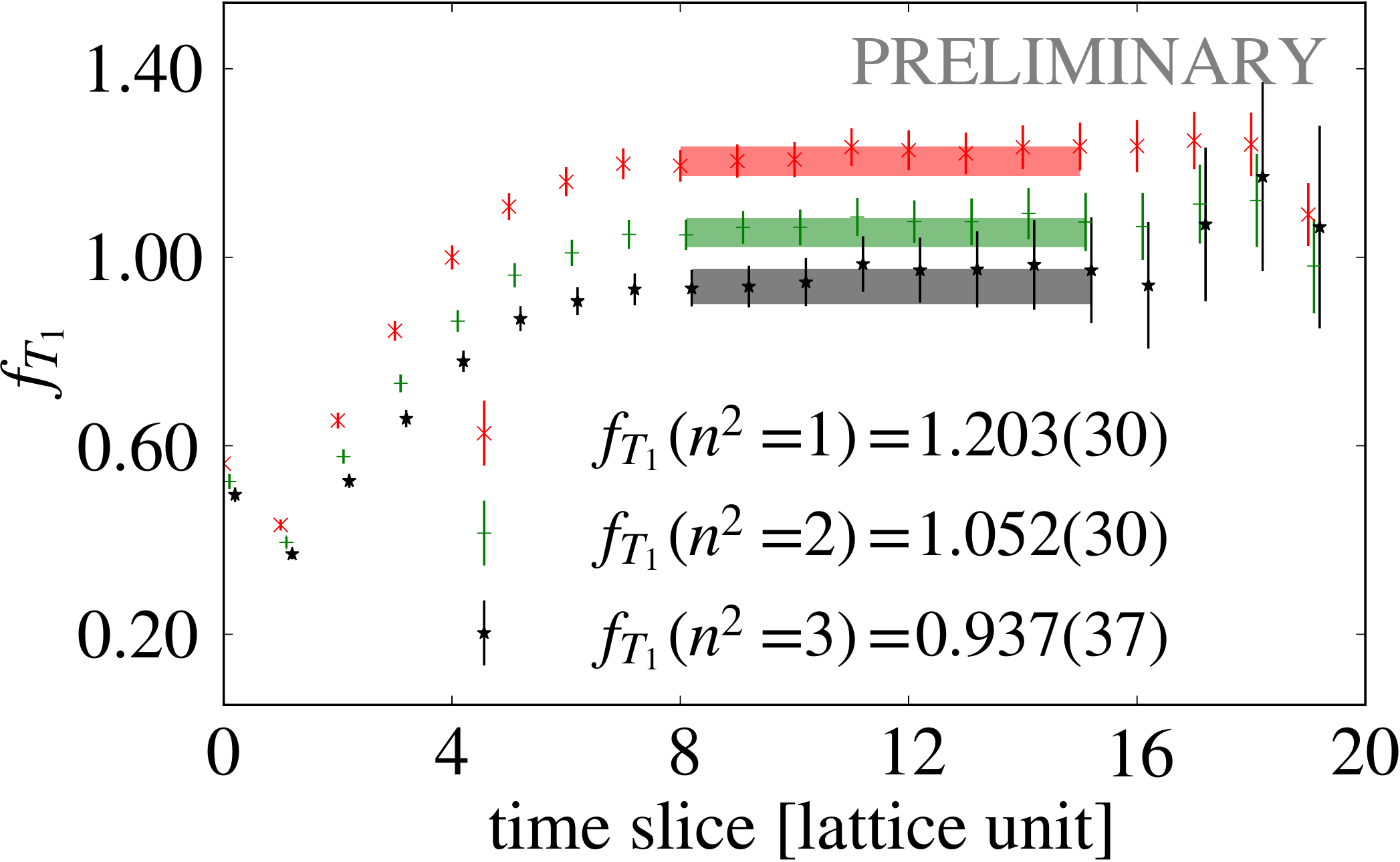}}}
  \put(0,20){
  \parbox{0.43\textwidth}{\includegraphics[height=0.20\textheight]{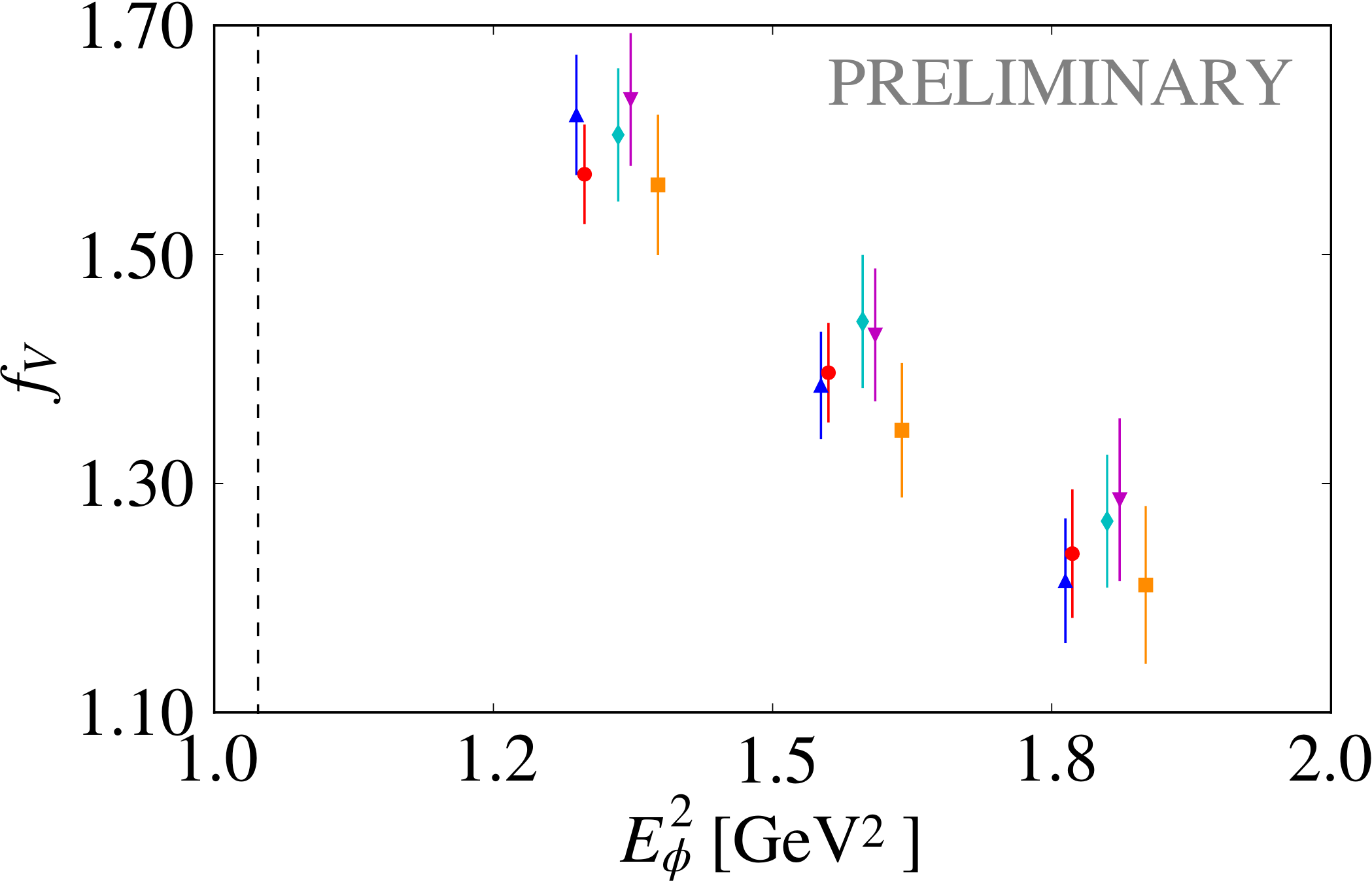}}\hspace{14.5mm}
  \parbox{0.43\textwidth}{\includegraphics[height=0.20\textheight]{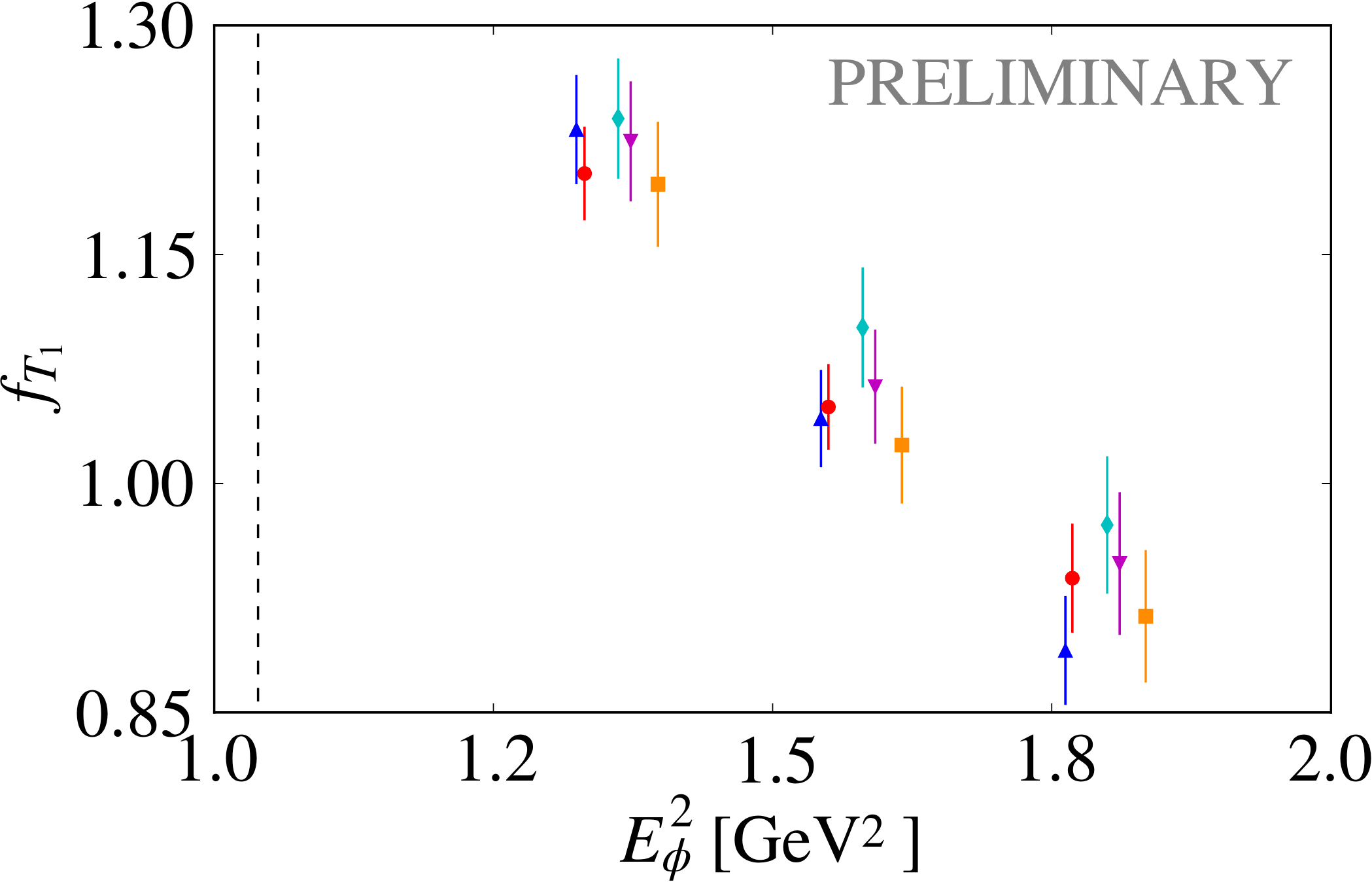}}}
  \put(15,7.2){\includegraphics[height=0.1\textheight]{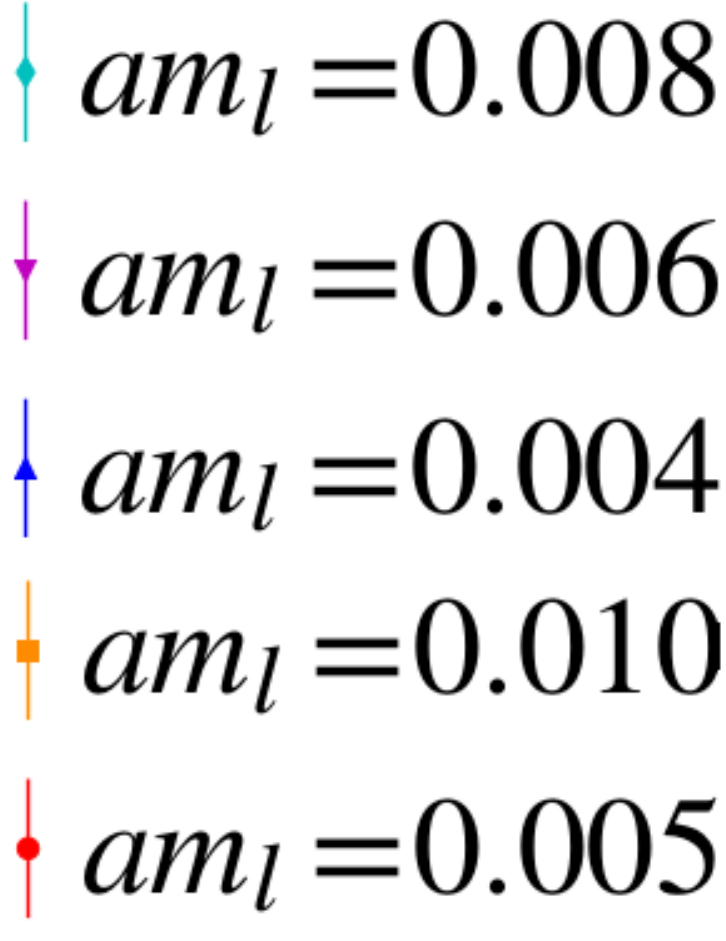}}
  \end{picture}
  \vspace{-1mm}
   \caption{Preliminary results for rare semi-leptonic form factors for $B_s\to\phi\ell^+\ell^-$ decays. Upper row: effective mass style plots for the form factors $f_V$ and $f_{T1}$; lower row dependence on the squared energy $E_\phi^2$. }
  \label{fig:RareDecays}
\end{figure}

Extending our program to calculate semi-leptonic decays, we generalized our code to compute form factors for decays forbidden at loop-level in the SM and/or resulting in a vector final state. In our current work \cite{Flynn:2015ynk}, we treat vector final states as stable and rely by that on the narrow width approximation. In the following we focus on rare decays mediated by a flavor changing neutral current (FCNC). Those decays are described in the SM by a basis of 20 operators contributing to the effective Hamiltonian \cite{Grinstein:1987vj,Grinstein:1990tj,Buras:1993xp,Ciuchini:1993ks,Ciuchini:1993fk,Ciuchini:1994xa}. Three of these operators, conventionally named ${\cal O}_7$, ${\cal O}_9$, and ${\cal O}_{10}$, are short distance dominated and suitable to be evaluated with current lattice QCD techniques. Possible implications due to charm resonances \cite{Lyon:2014hpa} remain to be addressed in future work. Operators  ${\cal O}_9$, and ${\cal O}_{10}$  correspond to the diagram sketched in Fig.~\ref{Fig:DiagramSketches}c;  ${\cal O}_7$ is depicted by Fig.~\ref{Fig:DiagramSketches}d. Like in the case of $B\to\pi\ell\nu$ semi-leptonic form factors, we need to compute 3-point functions to determine the short distance contributions which lead to a set of seven form factors: $f_V,\, f_{A_0},\, f_{A_1},\, f_{A_2},\, f_{T_1},\, f_{T_2},$ and $f_{T_3}$. As an example, we show in Fig.~\ref{fig:RareDecays} our preliminary results for the form factors $f_{V}$ and $f_{T_1}$ for case of the rare decay $B_s\to\phi\ell^+\ell^-$. The upper plots show our signal for the form factors vs.~the time slices between the $B_s$-meson located at $t_{sink}=20$ and the $\phi$-meson located at $t_0=0$. As before we keep the $B_s$-meson at rest and inject units on momentum on the side of the final state. For the shown form factors,  $f_{V}$ and $f_{T_1}$, only terms at nonzero momentum contribute and we find long and clean plateaus. In the lower plots we show the dependence on the squared energy as we resolve it using data on five ensembles at two different lattice spacings. Work is in progress to finalize our analysis, carry-out a chiral- and continuum extrapolation, and estimate systematic errors \cite{Flynn:2016vej}.

\paragraph{Acknowledgments} The authors thank our collaborators in
the RBC and UKQCD Collaborations for helpful discussions and suggestions.
Computations for this work were performed on resources provided by the USQCD
Collaboration, funded by the Office of Science of the U.S. Department of
Energy, as well as on computers at Columbia University and Brookhaven National
Laboratory. Gauge field configurations on which our calculations are based were
also generated using the DiRAC Blue Gene Q system at the University of
Edinburgh, part of the DiRAC Facility; funded by BIS National E-infrastructure
grant ST/K000411/1 and STFC grants ST/H008845/1, ST/K005804/1 and ST/K005790/1.
This project has received funding from the European Union's Horizon 2020 research
and innovation programme under the Marie Sk{\l}odowska-Curie grant agreement No 659322,
the European Research Council under the European Unions Seventh Framework Programme
(FP7/2007-2013) / ERC Grant agreement 279757, STFC grant ST/L000296/1 and
ST/L000458/1 as well as the EPSRC Doctoral Training Centre grant
(EP/G03690X/1).

{\small \bibliography{../B_meson} \bibliographystyle{apsrev4-1ow} }
        

\end{document}